\documentclass[%
superscriptaddress,
preprint,
longbibliography,
preprintnumbers,
 amsmath,amssymb,
 aps,
 prd,
]{revtex4-2}
\usepackage{CJK}
\usepackage{hyperref}
\usepackage{color}
\usepackage{graphicx}
\usepackage{xspace}
\usepackage{booktabs}
\usepackage{caption}
\usepackage{subcaption}

\newcommand{\pp}{\rm{pp}}

\newcommand{\Pp}{\rm{p}}
\newcommand{\Pe}{\rm{e}}
\newcommand{\PW}{\rm{W}}
\newcommand{\PV}{\rm{V}}
\newcommand{\PH}{\rm{H}}
\newcommand{\PZ}{\rm{Z}}
\newcommand{\TeV}{\rm{~TeV}}
\newcommand{\MGMCatNLO}{\textsc{MadGraph5}\_aMC@NLO}
\newcommand{\whizard}{\textsc{Whizard}}
\newcommand{\pythia}{\textsc{Pythia}}
\newcommand{\delphes}{\textsc{Delphes}}
\newcommand{\valencia}{\textsc{Valencia}}
\newcommand{\fastjet}{\textsc{FastJet}}
\newcommand{\pt}{p_{\mathrm{T}}\xspace}

\usepackage{lineno}

\begin{document}
\begin{CJK*}{GB}{} 

\title{Anomalous production of massive gauge boson pairs at muon colliders}

\author{Brad Abbott}
\affiliation{Homer L. Dodge Department of Physics and Astronomy, University of Oklahoma, Norman OK, USA}
\author{Aram Apyan}
\email{arapyan@brandeis.edu}
\affiliation{Department of Physics, Brandeis University, Waltham MA, USA}
\author{Bianca Azartash-Namin}
\affiliation{Homer L. Dodge Department of Physics and Astronomy, University of Oklahoma, Norman OK, USA}
\author{Veena Balakrishnan}
\affiliation{Homer L. Dodge Department of Physics and Astronomy, University of Oklahoma, Norman OK, USA}
\author{Jeffrey Berryhill}
\affiliation{Fermi National Accelerator Laboratory, Batavia IL, USA}
\author{Shih-Chieh Hsu}
\affiliation{Department of Physics, University of Washington, Seattle WA, USA}
\author{Sergo Jindariani}
\affiliation{Fermi National Accelerator Laboratory, Batavia IL, USA}
\author{Mayuri Prabhakar Kawale}
\affiliation{Homer L. Dodge Department of Physics and Astronomy, University of Oklahoma, Norman OK, USA}
\author{Elham E Khoda}
\affiliation{Department of Physics, University of Washington, Seattle WA, USA}
\author{Ryan Parsons}
\affiliation{Homer L. Dodge Department of Physics and Astronomy, University of Oklahoma, Norman OK, USA}
\author{Alexander Schuy}
\affiliation{Department of Physics, University of Washington, Seattle WA, USA}
\author{Michael Strauss}
\affiliation{Homer L. Dodge Department of Physics and Astronomy, University of Oklahoma, Norman OK, USA}
\author{John Stupak}
\affiliation{Homer L. Dodge Department of Physics and Astronomy, University of Oklahoma, Norman OK, USA}
\author{Connor Waits}
\affiliation{Homer L. Dodge Department of Physics and Astronomy, University of Oklahoma, Norman OK, USA}

\date{\today}

\begin{abstract}
The prospects of searches for anomalous production of hadronically decaying weak boson pairs at proposed high-energy muon colliders are reported. Muon-muon collision events are simulated at $\sqrt{s}=6$, 10, and 30 TeV, corresponding to an integrated luminosity of $4$, $10$, and $10$ ab$^{-1}$, respectively. Simulated $\mu\mu\rightarrow\PW\PW+\nu\nu/\mu\mu$ events are used to set expected constraints on the structure of quartic vector boson interactions in the framework of a dimension-8 effective field theory.  Similarly,  $\mu\mu\rightarrow\PW\PW/\PZ\PZ+\nu\nu$ events are used to report constraints on the product of the cross section and branching fraction for vector boson fusion production of a heavy neutral Higgs boson decaying to weak boson pairs. These results are interpreted in the context of the Georgi--Machacek model.
\end{abstract}


\maketitle
\end{CJK*}

 \newpage 


\section{Introduction}
\label{S:intro}

Vector boson scattering (VBS) processes probe the structure of the triple and quartic electroweak (EW) gauge boson self-interactions~\cite{Lee:1977yc,Lee:1977eg}. Deviations of measurements with respect to the Standard Model (SM) predictions could indicate the presence of anomalous quartic gauge couplings (aQGCs)~\cite{aqgc_operators, Almeida:2020ylr}. Measurements of VBS processes provide a unique insight into the EW symmetry breaking mechanism as the unitarity of the tree-level amplitude of the longitudinally polarized VBS at high energies is restored by a Higgs boson~\cite{Lee:1977yc,Lee:1977eg}. New physics models predict enhancements in VBS processes through extended Higgs sectors or modifications of the Higgs boson couplings to $\PW$ and $\PZ$ bosons~\cite{Espriu:2012ih,Chang:2013aya}.

A multi-TeV muon collider ($\mu^{+}\mu^{-}$)~\cite{Accettura:2023ked} is a ``high-luminosity weak boson collider''~\cite{Costantini:2020stv} and provides a great opportunity to study VBS processes. A comprehensive physics case for a future high-energy muon collider, with center of mass energies from 1 to 100 TeV, is reported in Ref.~\cite{AlAli:2021let}. A muon collider has considerable advantages compared to proposed linear and circular electron-positron ($e^{+}e^{-}$)~\cite{Bambade:2019fyw,CLICdp:2018cto,FCC:2018evy,CEPCStudyGroup:2018ghi} with larger collision energy and luminosity reach. In addition, a muon collider has a relatively clean environment compared to circular proton-proton ($\pp$) machines~\cite{FCC:2018vvp,CEPCStudyGroup:2018rmc} and the total energy of the muon is available in a collision in contrast to the dissociation of the composite proton. However, compared to $e^{+}e^{-}$ colliders, the effects of backgrounds induced by the muon beam decays, referred to as ``beam-induced background,'' are important and need to be studied in detail~\cite{Bartosik:2020xwr}. 

\begin{figure}[htb]
\centering
\includegraphics[width=0.49\textwidth]{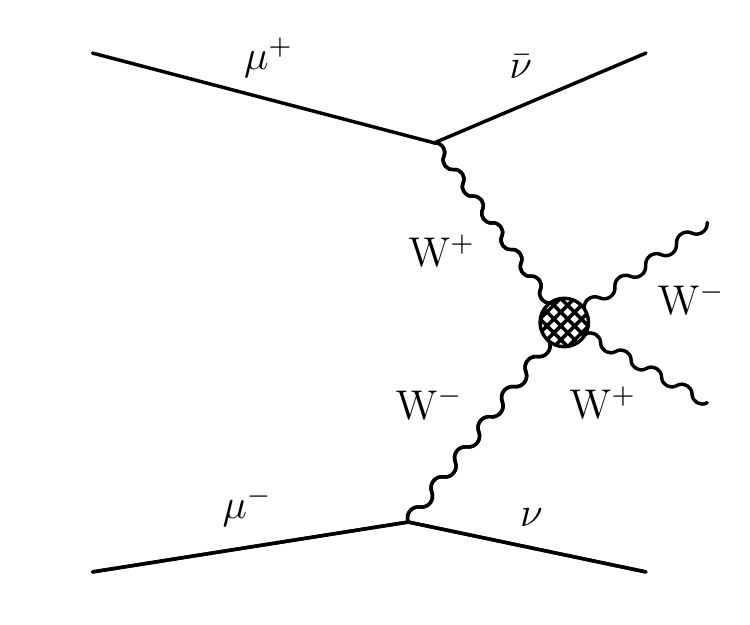}
\includegraphics[width=0.49\textwidth]{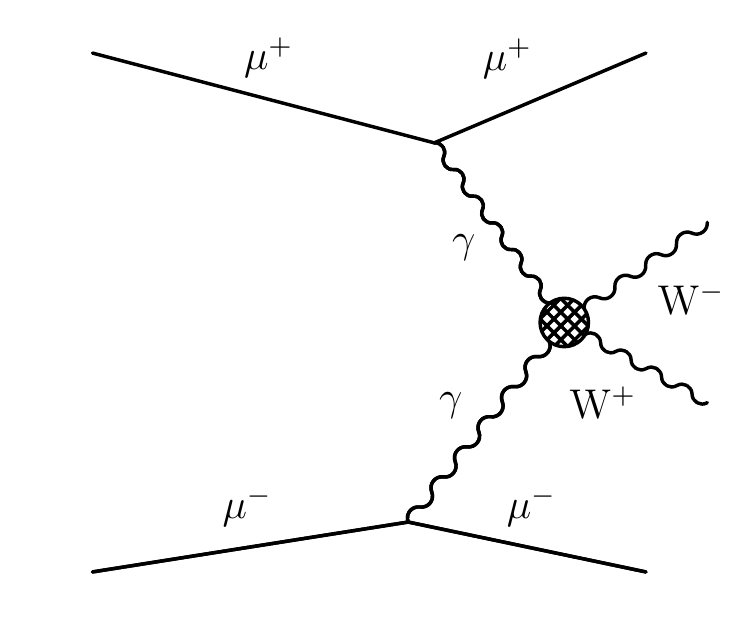}
\caption{Representative Feynman diagrams of the $\PW\PW\nu\nu$ (left) and $\PW\PW\mu\mu$ (right) processes. New physics (represented by a hatched circle) in the EW sector can modify the quartic gauge couplings.\label{fig:feynman}}
\end{figure}

This paper focuses on the prospects of aQGC searches using events with hadronically decaying $\PW^{\pm}\PW^{\mp}$ boson pairs. The studies are performed in the $\mu\mu\rightarrow\PW\PW+\nu\nu/\mu\mu$ channels, where the $\PW$ boson pair is produced in association with two neutrinos or two muons, respectively. Figure~\ref{fig:feynman} shows representative Feynman diagrams involving quartic vertices for the $\PW\PW\nu\nu$ (left) and $\PW\PW\mu\mu$ (right) channels. Ten independent charge conjugate and parity conserving dimension-8 effective operators are considered~\cite{aqgc_operators}. The \texttt{S0} and \texttt{S1} operators are constructed from the covariant derivative of the Higgs doublet. The \texttt{T0}, \texttt{T1}, \texttt{T2}, \texttt{T6}, and \texttt{T7} operators are constructed from the SU$_\mathrm{L}$(2) gauge fields. The mixed operators \texttt{M0}, \texttt{M1}, and \texttt{M7} involve the SU$_\mathrm{L}$(2) gauge fields and the Higgs doublet. The definitions of all the operators are provided in Appendix A in Ref.~\cite{aqgc_operators}. The $\PW\PW\nu\nu$ and $\PW\PW\mu\mu$ channels are analyzed separately.

Prospects of searches for a heavy neutral Higgs boson produced in association with two neutrinos and decaying to $\PW\PW$ or $\PZ\PZ$ boson pairs are also reported in this paper.  In particular, the Georgi--Machacek (GM) model~\cite{GEORGI1985463, CE1, CE2} with both real and complex triplets is considered. In the GM model the tree-level ratio of the $\PW$ and $\PZ$ boson masses is protected against large radiative corrections and the physical scalar states transform as multiplets under a global custodial symmetry. The model contains a fermiophobic fiveplet, a fermiophilic triplet, and two singlets, one of which is identified as the 125 GeV SM-like Higgs boson.

The fiveplet physical states, collectively referred to as $\PH_5$, consist of a neutral Higgs boson in addition to singly and doubly charged Higgs bosons, which are degenerate in mass (denoted as $m_{\PH_{5}}$).  Only the neutral fiveplet state is considered here. The \texttt{H5plane} benchmark,  where the triplet states are heavier than the fiveplet states, is used~\cite{LHCHiggsCrossSectionWorkingGroup:2016ypw}. In this benchmark, the  $\PH_{5}$ states are produced primarily via vector boson fusion (VBF) and the production cross section is proportional to the parameter $s_{\mathrm{H}}$, characterizing the contribution of the isotriplet  scalar fields to the masses of the W and Z bosons. The $\mu\mu\rightarrow\PH_{5}\nu\nu$ production mode is considered where the $\PH_{5}$ boson is assumed to decay to $\PV\PV$, where $\PV$ is a $\PW$ or $\PZ$ boson~\cite{Zaro:2002500,LHCHiggsCrossSectionWorkingGroup:2016ypw}, yielding a $\PV\PV\nu\nu$ final state.

Measurements of VBS processes at the CERN LHC by the ATLAS and CMS Collaborations have reported constraints on aQGCs in the framework of dimension-8 effective field theory (EFT) operators~\cite{ATLAS_ssWW,CMS_ssWW,Sirunyan:2017ret,Aaboud:2016ffv,Aad:2016ett,Sirunyan:2019ksz,Sirunyan:2017fvv,Khachatryan:2016mud,Khachatryan:2017jub,Khachatryan:2016vif,Aaboud:2017pds,Aaboud:2016uuk,Sirunyan:2019der,CMS:2020gfh,CMS:2020fqz, CMS:2020ypo,CMS:2021gme}.  Prospects for aQGC searches using the scattering of W and Z bosons at the High-Luminosity LHC (HL-LHC) and High-Energy LHC (HE-LHC) are reported in Ref.~\cite{Dainese:2703572}, while the sensitivity for a future $e^{+}e^{-}$ collider is presented in Ref.~\cite{Fleper:2016frz}. Constraints on the GM model have been reported by the ATLAS and CMS Collaborations by searching for charged Higgs bosons produced via VBF~\cite{PhysRevLett.114.231801,Sirunyan:2017sbn,Sirunyan:2017ret,Sirunyan:2019ksz,Sirunyan:2019der,CMS:2021wlt,ATLAS:2022zuc}.  

In this paper, $\mu^{+}\mu^{-}$ collider benchmarks~\cite{AlAli:2021let} with three different center of mass energies, $\sqrt{s}= \{6, 10, 30\}$ TeV, and integrated luminosities of $\{4, 10, 10 \}$ ab$^{-1}$, respectively, are considered. Events are selected with hadronically decaying $\PW$ or $\PZ$ bosons to target the final states with the highest branching ratios. Expected limits on aQGC parameters and constraints on the GM model are reported.

\section{Event simulation}
\label{S:simulation}

The \MGMCatNLO~3.1.1~\cite{Frederix2012,Alwall:2014hca} and \whizard~3~\cite{Moretti:2001zz,Kilian:2007gr} Monte Carlo (MC) event generators are used to simulate the signal and background contributions. The aQGC processes are simulated using \MGMCatNLO~at leading order (LO). The  contributions of the amplitude of the interference between the EFT operators and the SM (referred to as the interference term) are simulated separately from the contributions involving only EFT operators (referred to as the quadratic term). More details can be found in Ref.~\cite{ATL-PHYS-PUB-2023-002}. The $\PH_{5}$ signal processes are simulated using \MGMCatNLO~at LO for the mass range from 0.5 TeV to 3 TeV using the \texttt{H5plane} benchmark. The $s_{\mathrm{H}}$ values are set to 0.5 for masses up to 0.8 TeV and 0.25 for higher masses to be compatible with present constraints~\cite{LHCHiggsCrossSectionWorkingGroup:2016ypw}. The SM $\PW\PW\nu\nu$ and $\PW\PW\mu\mu$ background processes are simulated with \MGMCatNLO. These SM processes are also simulated with \whizard~at LO and good agreement is seen with \MGMCatNLO~predictions. 

Other background processes contributing to the $\PW\PW\nu\nu$ channel are simulated following Ref.~\cite{Fleper:2016frz}. The $\PW\PZ\mu\nu$, $\PZ\PZ\mu\mu$, $\PW\PW\mu\mu$, and $\PW\PW\PZ(\rightarrow\!\nu\nu)$ processes are simulated using \whizard. The initial state radiation of beam particles as implemented in \whizard~is included in the simulation.

The parton showering and hadronization are simulated using $\pythia$~8.306~\cite{Sjostrand:2014zea}. Detector effects are simulated using \delphes~3.5~\cite{deFavereau:2013fsa} with a generic muon collider detector description. The effects of beam-induced background are not considered in this description. Muons and electrons are reconstructed with an absolute pseudorapidity up to 2.5. Jets are clustered from the reconstructed stable particles (except electrons and muons) using \fastjet~\cite{Cacciari:2011ma} with the \valencia~algorithm~\cite{Boronat:2014hva}. Inclusive clustering with a distance parameter of $R=1$ is performed. 

\section{Event selection}
\label{S:sel}

Events are selected targeting hadronically decaying $\PW\PW$ and $\PZ\PZ$ boson pairs with a large invariant mass. The jets are required to have transverse momenta ($\pt$) greater than 100 GeV and be relatively central, with $|\cos \theta| < 0.8$, where $\theta$ is the angle of the jet with respect to the beam axis. The jet with the highest $\pt$ is called the ``leading jet'' and the jet with the second-highest $\pt$ the ``subleading jet.'' The leading and subleading jets are each required to have a mass greater than 40 GeV.


The $\PW\PW\nu\nu$ and $\PZ\PZ\nu\nu$ channels are targeted by vetoing events with a reconstructed electron or muon with momentum greater than $3$ GeV. This requirement significantly reduces the $\PW\PW\mu\mu$ and $\PZ\PZ\mu\mu$ background contributions in these channels, which could be reduced even further with a detector with better forward muon coverage. As these channels contain two neutrinos in the final state, the events are also required to have a missing mass ($m_\mathrm{miss}$) greater than $200$ GeV. The $m_\mathrm{miss}$ is defined as   
\begin{equation}
m_\mathrm{miss} = \sqrt{(\sqrt{s}-E_{\PV\PV})^2-|\vec{p}_{\PV\PV}|^{2}},
\label{eq:missingmass}
\end{equation}
where $E_{\PV\PV}$ and $\vec{p}_{\PV\PV}$ are the energy and momentum of the $\PV$ boson pair. This requirement removes events where neutrinos are produced from $\PZ$ boson decays and reduces the contributions of the s-channel $\PW\PW$ and two-jet processes.

The $\PW\PW\mu\mu$ channel is targeted by requiring two oppositely charged muons with momenta greater than $0.5$ GeV and $|\cos \theta| < 0.99$. The largely dominant background contribution is the SM production of $\PW\PW\mu\mu$ where the final state muons tend to be very forward. The mass of the dimuon pair is required to be greater than 106 GeV to reduce the contribution of events where the muons are produced from $\PZ$ boson decays.

\section{Results}
\label{S:results}

\begin{figure}[htbp]
\centering
\includegraphics[width=0.38\textwidth]{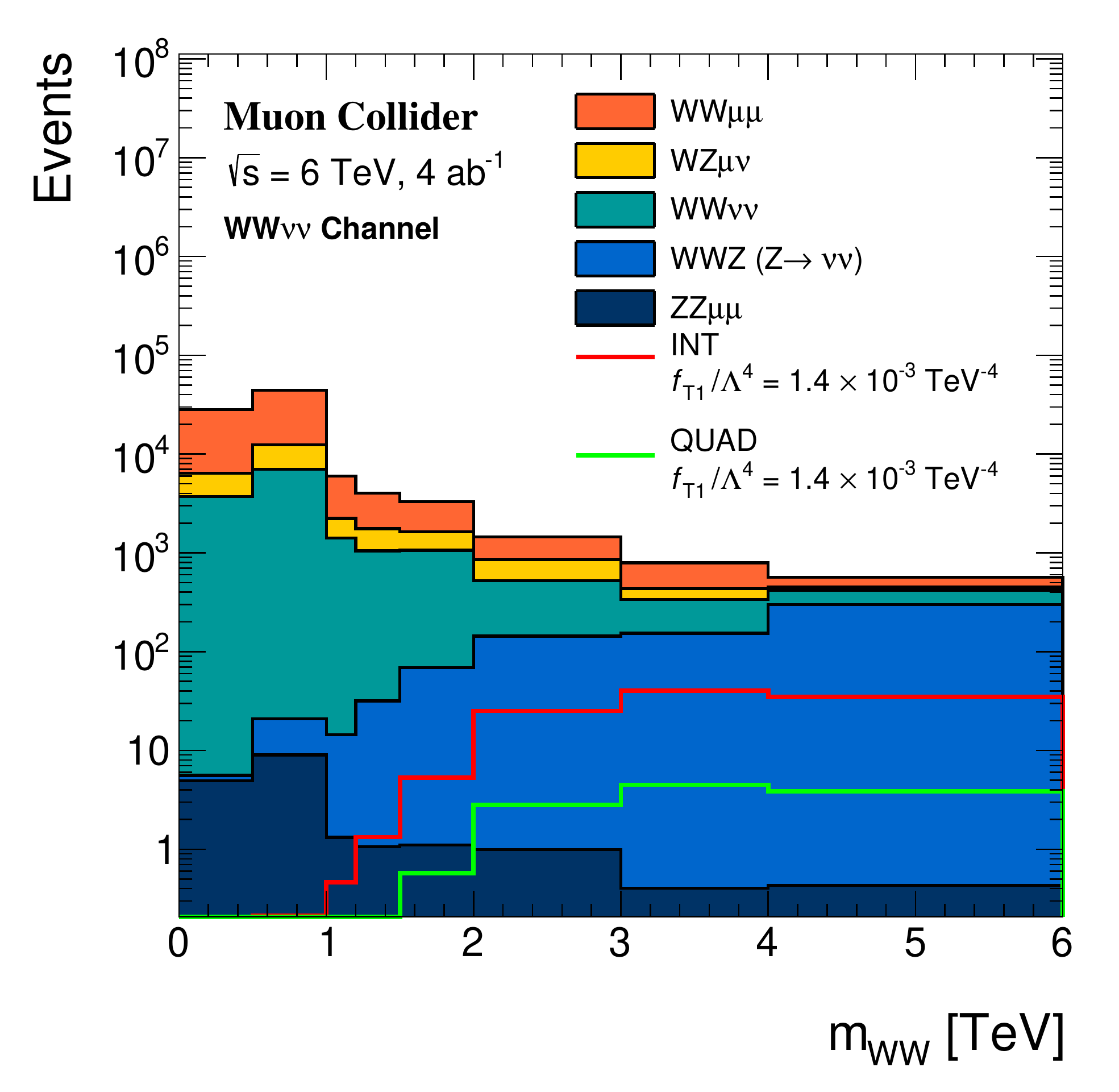} 
\includegraphics[width=0.38\textwidth]{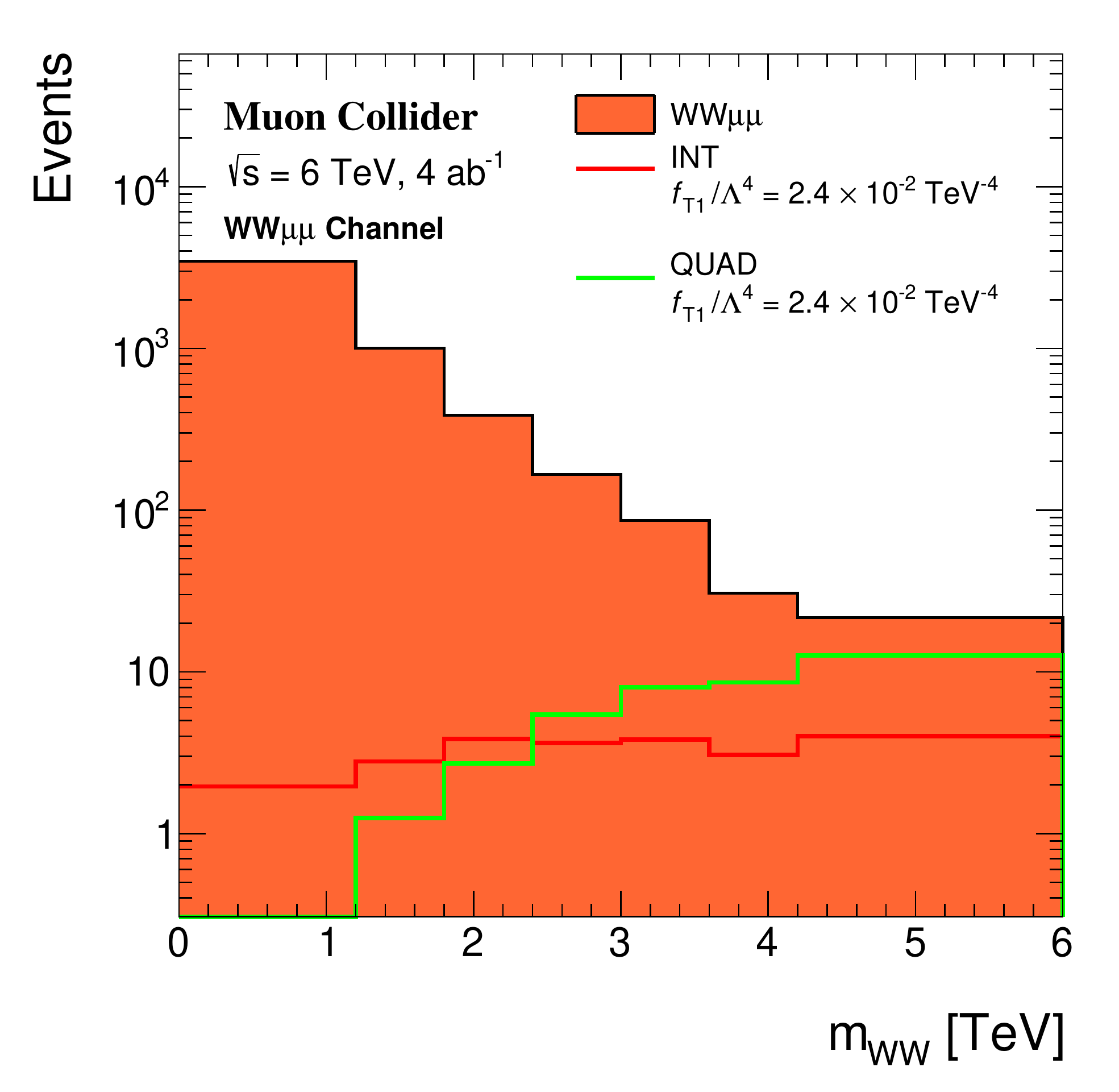}
\includegraphics[width=0.38\textwidth]{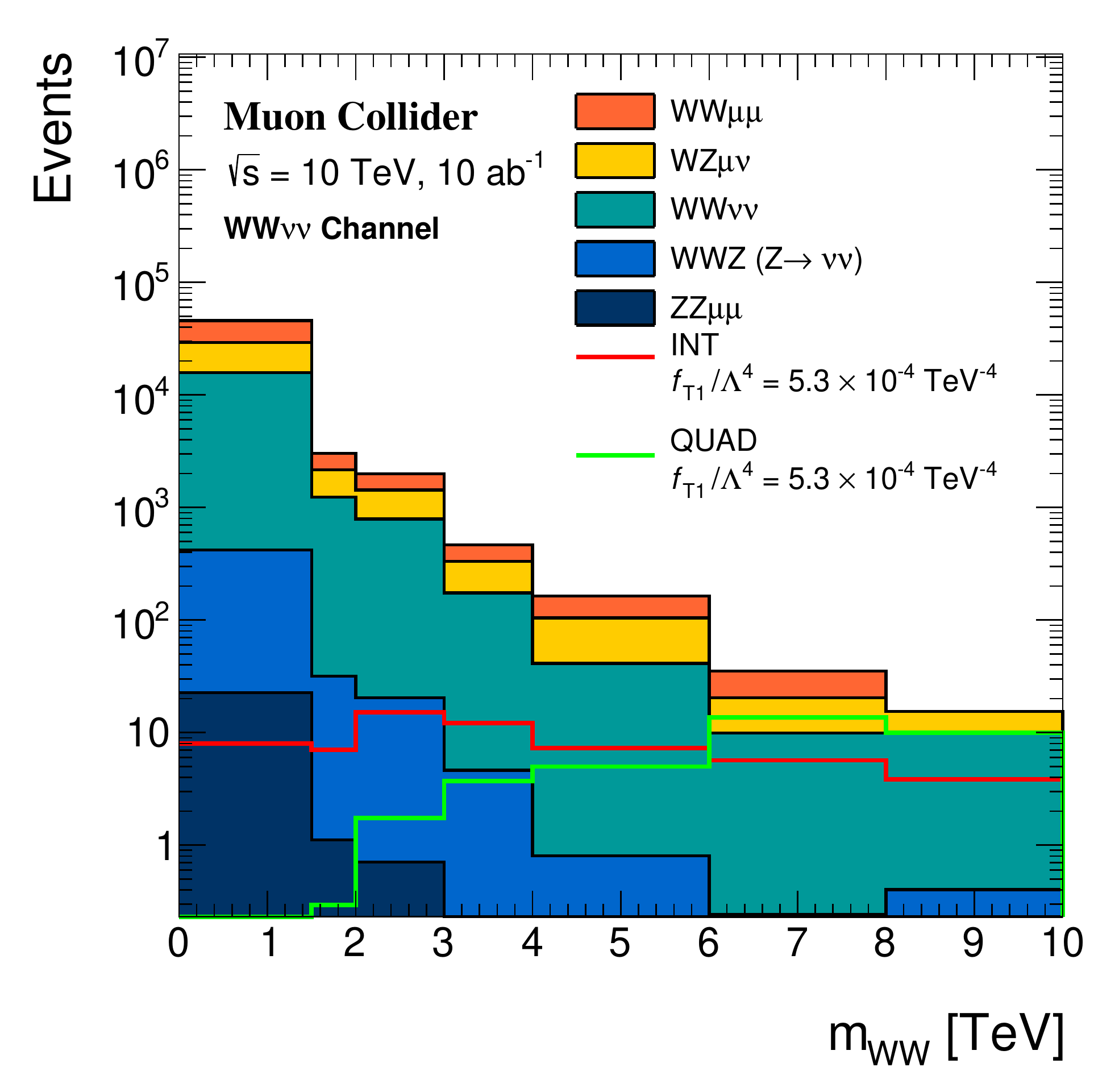} 
\includegraphics[width=0.38\textwidth]{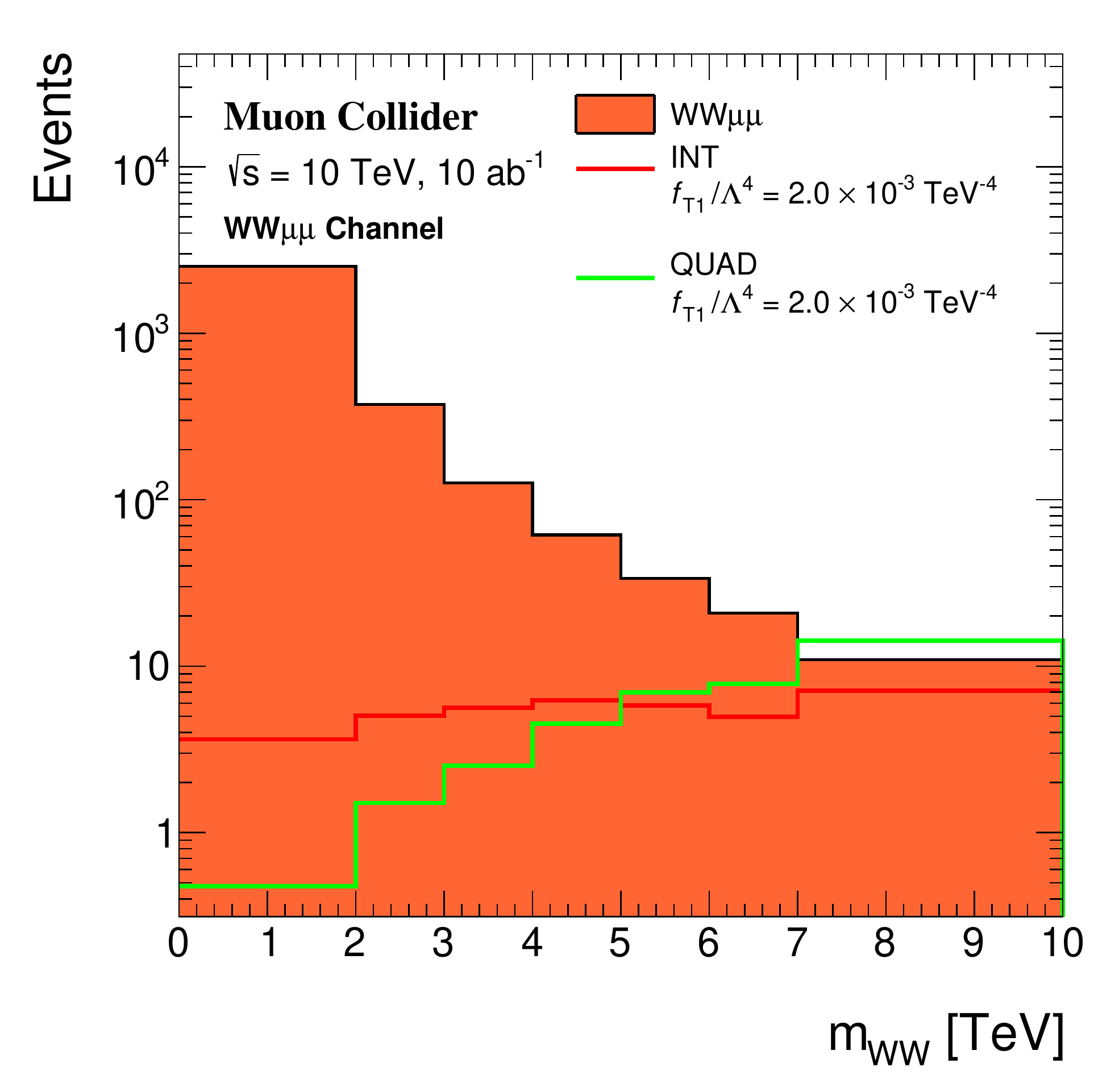}
\includegraphics[width=0.38\textwidth]{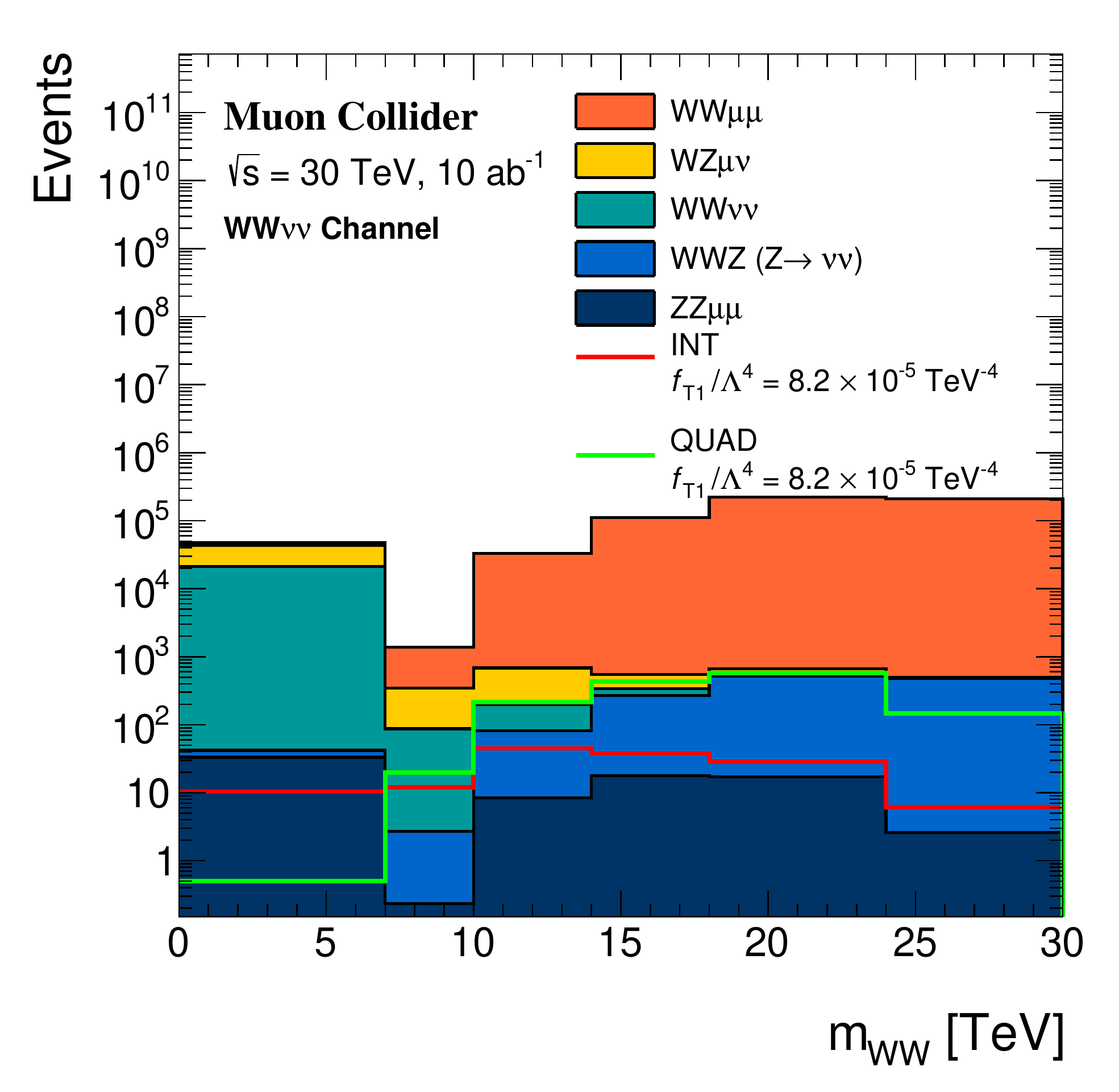} 
\includegraphics[width=0.38\textwidth]{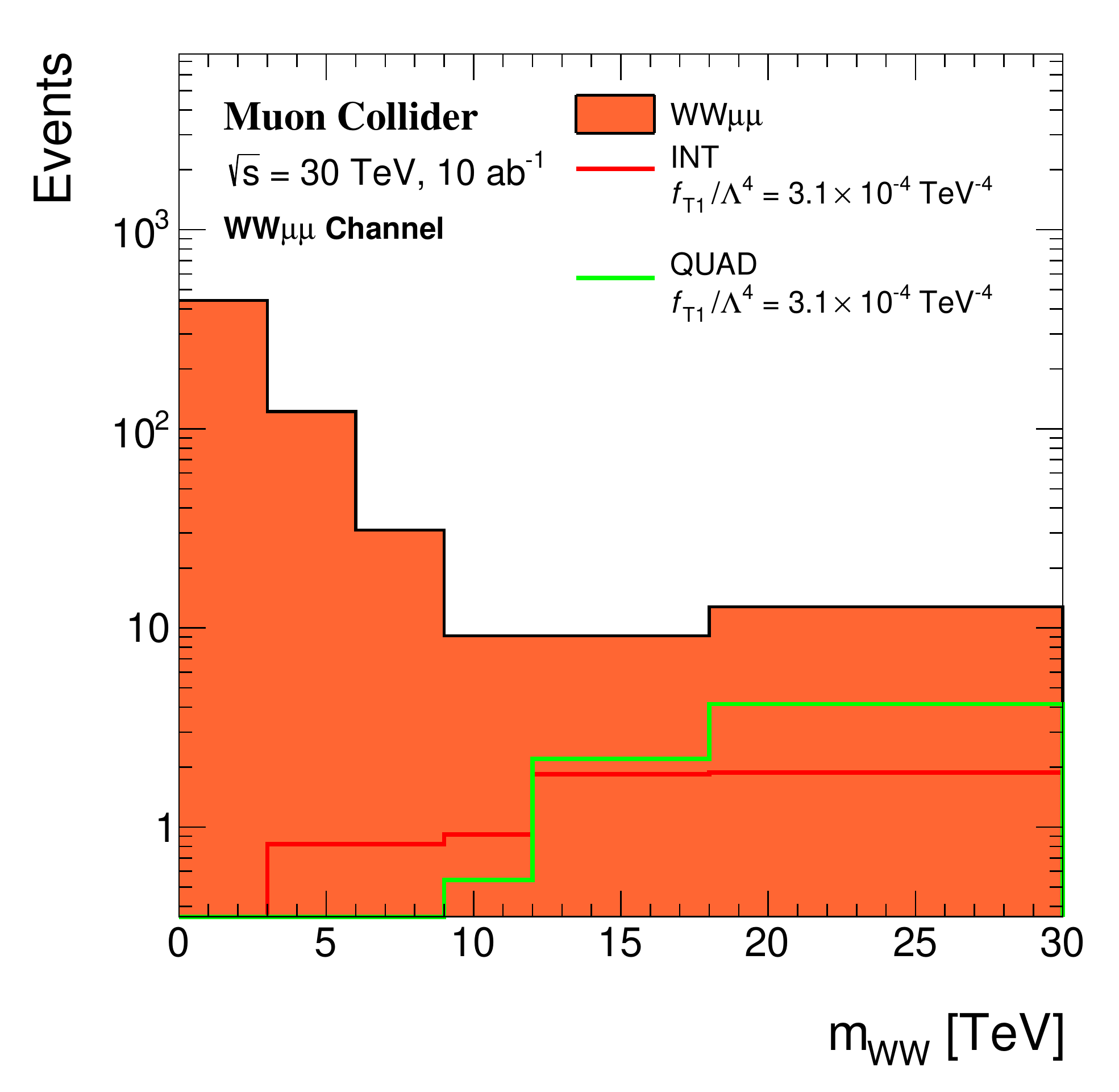}
\caption{Distributions of $m_{\PW\PW}$ after the event selection in the $\PW\PW\nu\nu$ (left) and $\PW\PW\mu\mu$ (right) channels for a $\mu^{+}\mu^{-}$ collider with $\sqrt{s}=6$ TeV (upper), 10 TeV (middle), and 30 TeV (lower). The filled histograms show the background expectation, while the solid lines show the separate contributions from the interference (red) and quadratic (green) terms for a  value of the \texttt{T1} parameter at the limit of the expected sensitivity. 
}
\label{fig:dijet_mass}
\end{figure}

The selected events are used to constrain aQGC parameters in an EFT framework. Statistical analysis of the event yields is performed separately in the $\PW\PW\nu\nu$ and $\PW\PW\mu\mu$ channels with a fit to the invariant mass distribution of the two leading jets, which typically correspond to the pair of $\PW$ bosons, denoted $m_{\PW\PW}$. The distributions of $m_{\PW\PW}$ after the event selection at each of the different center-of-mass energies are shown in Fig.~\ref{fig:dijet_mass}. The expected $95\%$ confidence level (CL) lower and upper limits on the aQGC parameters $f/\Lambda^4$, where $f$ is the Wilson coefficient of the given operator and $\Lambda$ is the energy scale of new physics, are derived from Wilk's theorem~\cite{wilks:10.1214} assuming that the profile likelihood test statistic is $\chi^2$ distributed~\cite{CLs}. No nuisance parameters corresponding to systematic uncertainties are included in the fits. 

Table~\ref{tab:limits_nunu} shows the individual lower and upper limits obtained by setting all other aQGC parameters to zero in the $\PW\PW\nu\nu$ channel for the \texttt{S0}, \texttt{S1}, \texttt{M0}, \texttt{M1}, \texttt{M7}, \texttt{T0}, \texttt{T1}, and \texttt{T2} operators, at each of the different center-of-mass energies. The $\PW\PW\mu\mu$ contribution in the $\PW\PW\nu\nu$ channel is treated as a background process and assumed to be purely SM in the statistical analysis. Table~\ref{tab:limits_mumu} shows the individual lower and upper limits obtained by setting all other aQGC parameters to zero in the $\PW\PW\mu\mu$ channel for the \texttt{T0}, \texttt{T1}, \texttt{T2}, \texttt{T6}, and \texttt{T7} operators for the different center-of-mass energies. The operators \texttt{T6} and \texttt{T7} are especially interesting for the $\PW\PW\mu\mu$ channel as the presence of these operators does not modify the SM quartic $\PW\PW\PW\PW$ vertex.

\begin{table}[htb]
\centering
\caption{Expected lower and upper 95\% CL limits on the parameters of the quartic operators \texttt{S0}, \texttt{S1}, \texttt{S2}, \texttt{M0}, \texttt{M1}, \texttt{M7}, \texttt{T0}, \texttt{T1}, and \texttt{T2} in the $\PW\PW\nu\nu$ channel for a $\mu^{+}\mu^{-}$ collider with $\sqrt{s}=6$ TeV, 10 TeV, and 30 TeV. The energy at which tree-level unitarity would be violated for these parameter values is also shown.}\label{tab:limits_nunu}
\vspace{0.3cm}
\scriptsize
\begin{tabular}{c|cc|cc|cc}
\toprule
$\PW\PW\nu\nu$ &\multicolumn{2}{c|}{$\sqrt{s}= 6$~TeV}&\multicolumn{2}{c|}{$\sqrt{s}= 10$~TeV} &\multicolumn{2}{c}{$\sqrt{s}= 30$~TeV}\\
 & Limit & Unitarity & Limit & Unitarity & Limit & Unitarity \\
 &  (TeV\(^{-4}\)) & Bound (TeV)  &  (TeV\(^{-4}\)) & Bound (TeV)  &  (TeV\(^{-4}\)) & Bound (TeV) \\
\midrule
  \(f_{\mathrm{M},0}/\Lambda^4\) & \([-0.025, 0.027]\) & \([5.9, 5.8]\) &    \([-0.0048, 0.0049]\) & \([8.9, 8.8]\)   &        \([-0.00046, 0.00046]\) & \([15.8, 15.8]\)   \\
 \(f_{\mathrm{M},1}/\Lambda^4\) &  \([-0.063, 0.052]\) & \([6.6, 6.9]\) &      \([-0.0096, 0.0084]\) & \([10.5, 10.8]\)  &    \([-0.0012, 0.0011]\) & \([17.6, 17.8]\)     \\ 
 \(f_{\mathrm{M},7}/\Lambda^4\) &  \([-0.094, 0.12]\) & \([7.1, 6.7]\) &       \([-0.016, 0.019]\) & \([10.9, 10.6]\)  &      \([-0.0021, 0.0022]\) & \([18.1, 17.9]\)     \\ 
 \(f_{\mathrm{S},0}/\Lambda^4\) &  \([-0.19, 0.18]\) & \([3.8, 4.4]\) &        \([-0.034, 0.033]\) & \([5.8, 6.8]\)  &        \([-0.0046, 0.0045]\) & \([9.5, 10.9]\)      \\ 
 \(f_{\mathrm{S},1}/\Lambda^4\) &  \([-0.11, 0.11]\) & \([4.5, 4.3]\) &        \([-0.019, 0.019]\) & \([6.8, 6.6]\)  &        \([-0.0025, 0.0025]\) & \([11.3, 10.9]\)     \\ 
 \(f_{\mathrm{S},2}/\Lambda^4\) &  \([-0.11, 0.11]\) & \([4.4, 4.3]\) &        \([-0.019, 0.019]\) & \([6.8,  6.6]\)  &       \([-0.0025, 0.0025]\) & \([11.3, 10.9]\)     \\ 
 \(f_{\mathrm{T},0}/\Lambda^4\) &  \([-0.0049, 0.0025]\) & \([6.2, 6.3]\) &    \([-0.00070, 0.00051]\) & \([10.0, 9.3]\)  &   \([-0.000072, 0.000062]\) & \([17.7, 15.7]\) \\  
 \(f_{\mathrm{T},1}/\Lambda^4\) &  \([-0.0017, 0.0014]\) & \([7.7, 8.1]\) &    \([-0.00089, 0.00053]\) & \([9.0, 10.3]\)  &   \([-0.000095, 0.000082]\) & \([15.5, 16.3]\) \\  
 \(f_{\mathrm{T},2}/\Lambda^4\) &  \([-0.011, 0.0046]\) & \([6.6, 7.0]\) &    \([-0.0015, 0.00082]\) & \([10.8, 10.7]\)  &    \([-0.00017, 0.00013]\) & \([18.4, 16.7]\)   \\ 
\bottomrule
\end{tabular}
\end{table}

\begin{table}[htb]
\centering
\caption{Expected lower and upper 95\% CL limits on the parameters of the quartic operators \texttt{T0}, \texttt{T1}, \texttt{T2}, \texttt{T6}, \texttt{T7} in the $\PW\PW\mu\mu$ channel for a $\mu^{+}\mu^{-}$ collider with $\sqrt{s}=6$ TeV, 10 TeV, and 30 TeV. The energy at which tree-level unitarity would be violated for these parameter values is also shown. \label{tab:limits_mumu}}
\vspace{0.3cm}
\scriptsize
\begin{tabular}{c|cc|cc|cc}
\toprule
$\PW\PW\mu\mu$ &\multicolumn{2}{c|}{$\sqrt{s}= 6$~TeV}&\multicolumn{2}{c|}{$\sqrt{s}= 10$~TeV} &\multicolumn{2}{c}{$\sqrt{s}= 30$~TeV}\\
 & Limit & Unitarity & Limit & Unitarity & Limit & Unitarity \\
 &  (TeV\(^{-4}\)) & Bound (TeV)  &  (TeV\(^{-4}\)) & Bound (TeV)  &  (TeV\(^{-4}\)) & Bound (TeV) \\
\midrule
\(f_{\mathrm{T},0}/\Lambda^4\) &  \([-0.0065, 0.0026]\) & \([8.7, 10.9]\)   & \([-0.0012, 0.00057]\) & \([13.2, 15.8]\)     & \([-0.000031, 0.000018]\) & \([32.2, 36.8]\) \\
\(f_{\mathrm{T},1}/\Lambda^4\) &  \([-0.036, 0.024]\) & \([4.2, 5.2]\)      & \([-0.0020, 0.00089]\) & \([8.6, 11.9]\)      & \([-0.000050, 0.000031]\) & \([21.1, 27.0]\) \\
\(f_{\mathrm{T},2}/\Lambda^4\) &  \([-0.052, 0.030]\) & \([5.2, 6.7]\)      & \([-0.0068, 0.0012]\) & \([8.6, 15.1]\)       & \([-0.000091, 0.000042]\) & \([24.8, 35.1]\) \\
\(f_{\mathrm{T},6}/\Lambda^4\) &  \([-0.0052, 0.0041]\) & \([10.0, 11.1]\)  & \([-0.00090, 0.00074]\) & \([15.5, 16.9]\)    & \([-0.000027, 0.000024]\) & \([37.1, 40.1]\) \\
\(f_{\mathrm{T},7}/\Lambda^4\) &  \([-0.0068, 0.0042]\) & \([12.8, 15.0]\)  & \([-0.0011, 0.00086]\) & \([19.9, 22.3]\)     & \([-0.000034, 0.000028]\) & \([47.8, 52.7]\) \\
\bottomrule
\end{tabular}
\end{table}

The EFT framework is not a complete model and the presence of nonzero aQGCs will violate tree-level unitarity at sufficiently high energy~\cite{Almeida:2020ylr}. The physicality of the obtained limits can be deduced by investigating the perturbative partial-wave unitarity. While detailed studies on the EFT framework validity are beyond the scope of this paper, the unitarity bounds were evaluated for each aQGC parameter limit by calculating the $\PV\PV$ center-of-mass energy at which the tree-level unitarity would be violated without a form factor using \textsc{vbfnlo} 1.4.0~\cite{Arnold:2008rz,Arnold:2011wj,Baglio:2014uba}. These unitarity bounds are shown in Tables~\ref{tab:limits_nunu} and~\ref{tab:limits_mumu}. Various $\PV\PV \to \PV\PV $ channel contributions to the zeroth partial wave are considered and the smallest unitarity bound is chosen.  Generally, for 6 and 10 TeV collider options, unitarity violation occurs around or above the collider center-of-mass energy. 
On the other hand, the expected limits at $\sqrt{s} = 30$~TeV are somewhat optimistic as the corresponding unitarity bounds are significantly smaller than $30$ TeV for many of the operators. 

These results give stringent constraints on the aQGC parameters for the \texttt{S0}, \texttt{S1}, \texttt{M0}, \texttt{M1}, \texttt{M6}, \texttt{M7}, \texttt{T0}, \texttt{T1}, \texttt{T2}, \texttt{T5}, and \texttt{T6} operators. Depending on the operator, the expected limits are better by more than one or two orders of magnitude compared to the expected limits at the HL-LHC and HE-LHC, as reported in Ref.~\cite{Dainese:2703572}, and summarized in Table~\ref{tab:LHC}. The expected limits in the $\PW\PZ$ channel are based on a measurement of fully leptonic WZ scattering by the ATLAS Collaboration using $\pp$ collisions at $\sqrt{s}=13$~TeV~\cite{Aaboud_2019} with additional cuts to enhance the sensitivity to new physics, while those in the $\PW^{\pm}\PW^{\pm}$ channel are based on simulated $\pp$ collisions with same-sign leptons at $\sqrt{s}=14$~TeV~\cite{ATL-PHYS-PUB-2018-052} with an upgraded ATLAS detector~\cite{ATL-PHYS-PUB-2019-005}. Results for the HE-LHC are obtained based on simulations at $\sqrt{s}=27$~TeV, assuming the same signal-to-background ratio as at the LHC. A few of these LHC results address the unitarity issues in some form, but not all of them. Expected sensitivity to aQGCs at a $\sqrt{s} = 30$~TeV $\mu^{+}\mu^{-}$ collider in the $\PW\PW\nu\nu$ channel using events with leptonically decaying $\PW$ bosons is reported in Ref.~\cite{Yang:2022fhw}.  Sensitivity to aQGCs at high-energy $e^{+}e^{-}$ colliders are reported in Refs.~\cite{Dainese:2703572,Fleper:2016frz}. 


\begin{table}[htbp]\caption{Summary of expected limits (in TeV$^{-4}$) on the parameters of quartic operators at the HL-LHC and HE-LHC~\cite{Dainese:2703572}.}

\begin{tabular}{|c|c|c|c|c|} \hline
& \multicolumn{2}{|c|}{HL-LHC} & \multicolumn{2}{|c|}{HE-LHC} \\
& $\PW\PZ$ & $\PW^{\pm}\PW^{\pm}$ & $\PW\PZ$ & $\PW^{\pm}\PW^{\pm}$ \\ \hline\hline
$f_{S0}/\Lambda^4$ & [$-8, 8$] & [$-6,6 $] & [$-1.5,1.5 $] & [$-1.5,1.5 $]\\
$f_{S1}/\Lambda^4$ & [$-18,18$] & [$-16,16 $] & [$-3,3 $] & [$-2.5,2.5 $]\\
$f_{T0}/\Lambda^4$ & [$-0.76,0.76$] & [$-0.6,0.6 $] & [$-0.04,0.04 $] & [$-0.027,0.027 $]\\
$f_{T1}/\Lambda^4$ & [$-0.50,0.50$] & [$-0.4,0.4 $] & [$-0.03,0.03 $] & [$-0.016,0.016 $]\\
$f_{M0}/\Lambda^4$ & [$-3.8, 3.8$] & [$-4.0,4.0 $] & [$-0.5,0.5 $] & [$-0.28,0.28 $]\\
$f_{M1}/\Lambda^4$ & [$-5.0,5.0$] & [$-12,12 $] & [$-0.8,0.8 $] & [$-0.90,0.90 $]\\ \hline
\end{tabular}
\label{tab:LHC}
\end{table}

\begin{figure}[htbp]
\centering
\includegraphics[width=0.38\textwidth]{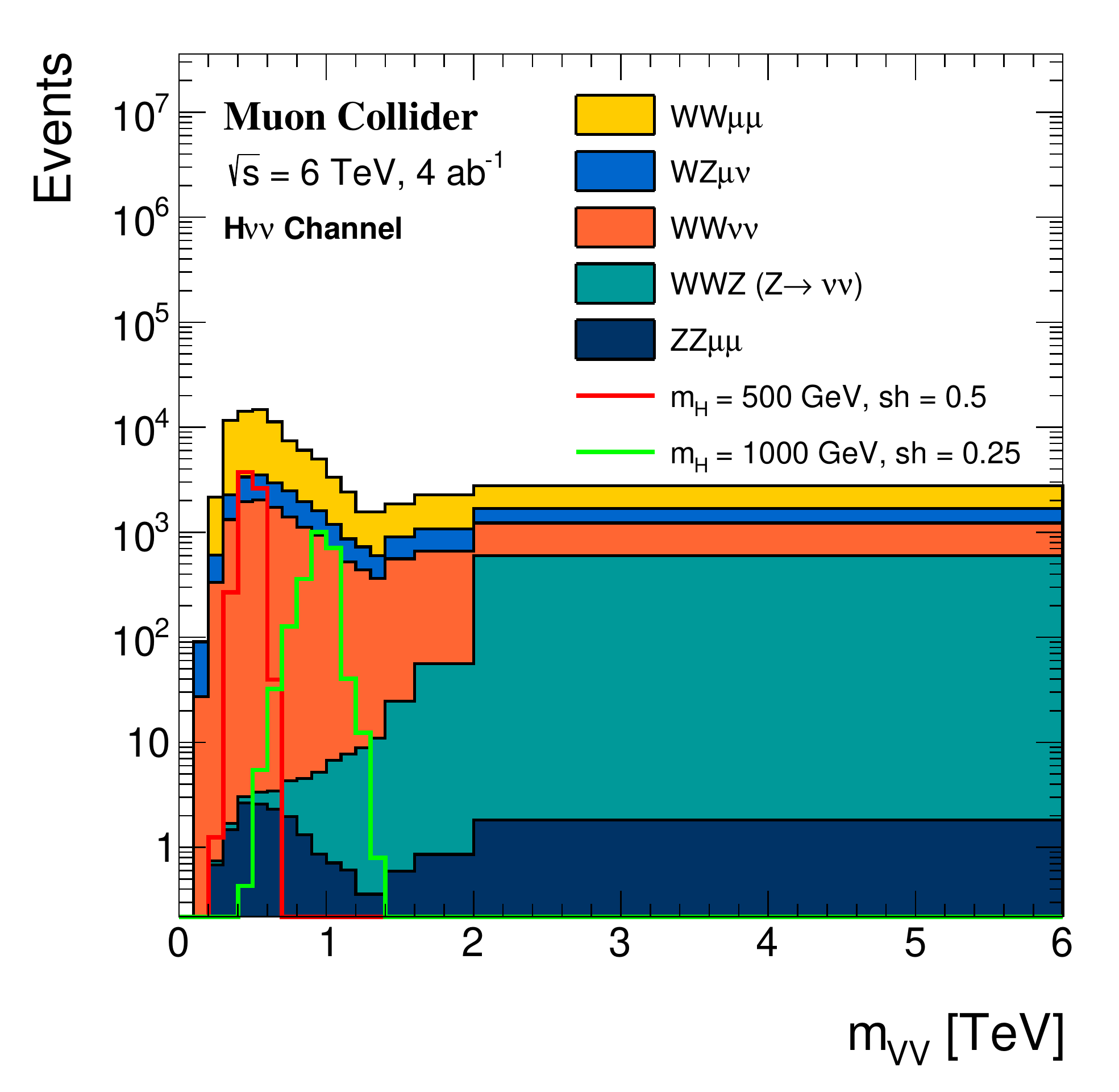} 
\includegraphics[width=0.38\textwidth]{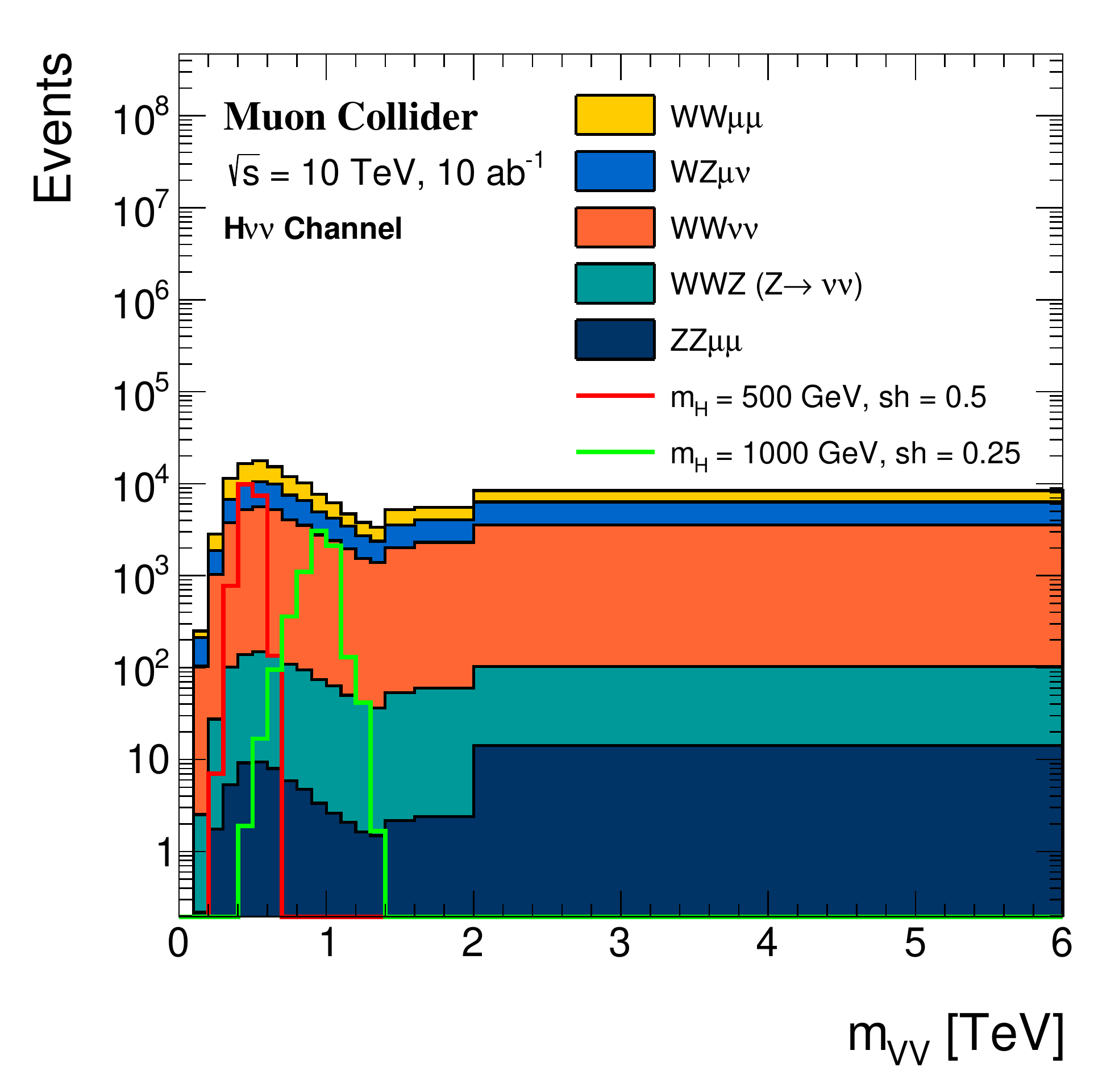} 
\includegraphics[width=0.38\textwidth]{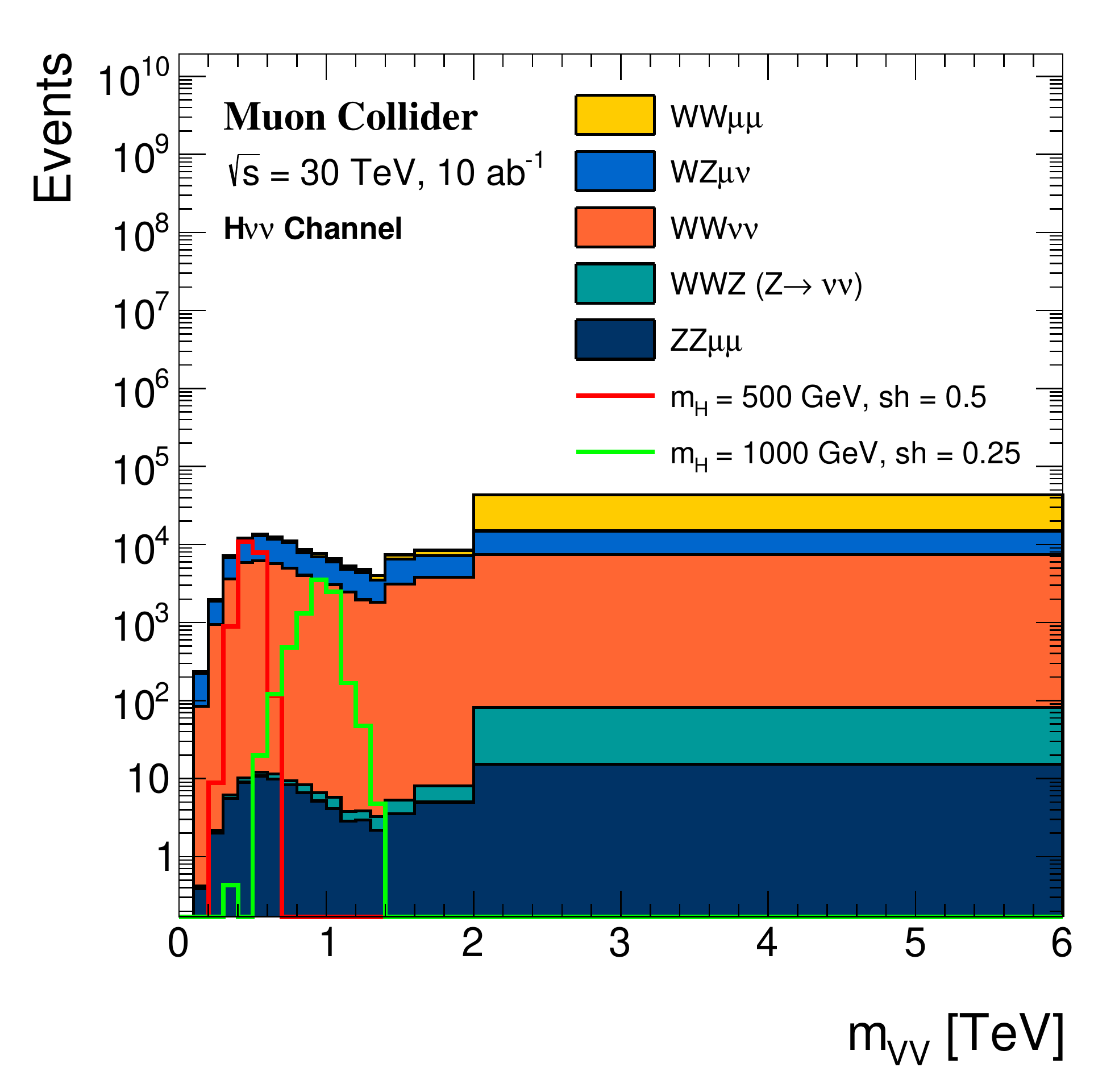}
\caption{Distributions of $m_{\PV\PV}$ after the event selection for a $\mu^{+}\mu^{-}$ collider with $\sqrt{s}=6$ TeV (upper left), 10 TeV (upper right), and 30 TeV (lower). The filled histograms show the background expectation, while the solid lines show the GM neutral Higgs signal predictions for values of $s_{\PH}=0.5$ and $m_{\PH_{5}}=500$ GeV (red), as well as for values $s_{\PH}=0.25$ and $m_{\PH_{5}}=1000$ GeV (green). Overflow is included in the last bin.\label{fig:dijet_mass_higgs}}
\end{figure}

\begin{figure}[htbp]
\centering
\includegraphics[width=0.45\textwidth]{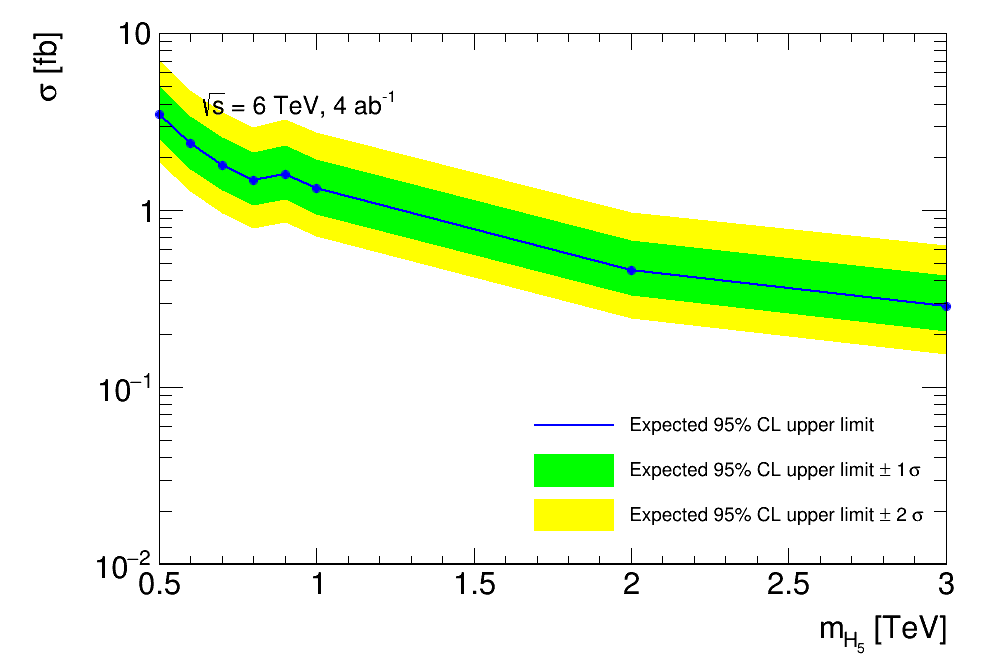} 
\includegraphics[width=0.45\textwidth]{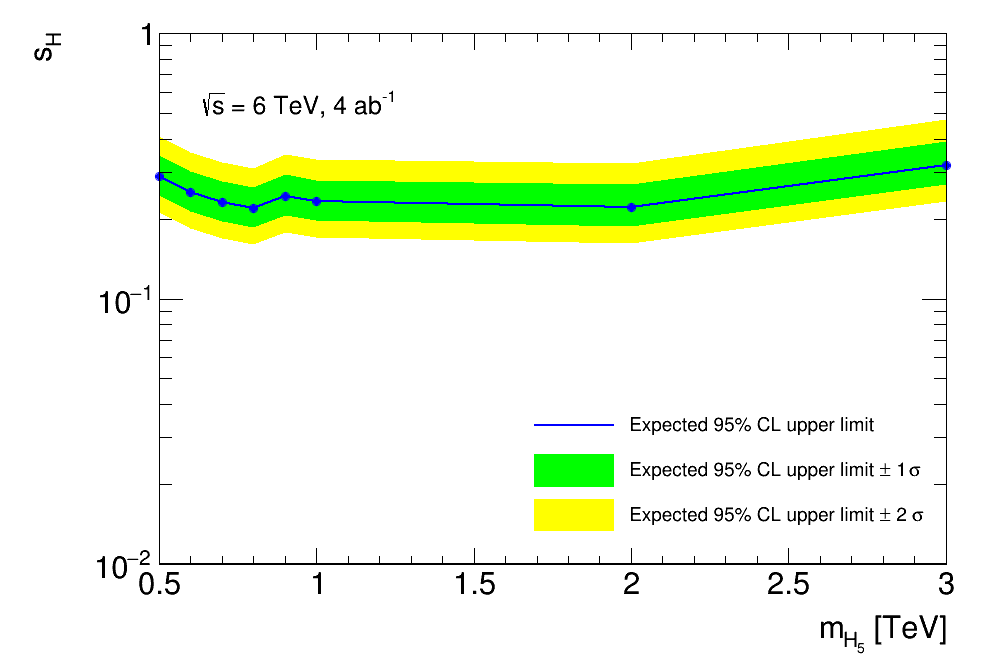}
\includegraphics[width=0.45\textwidth]{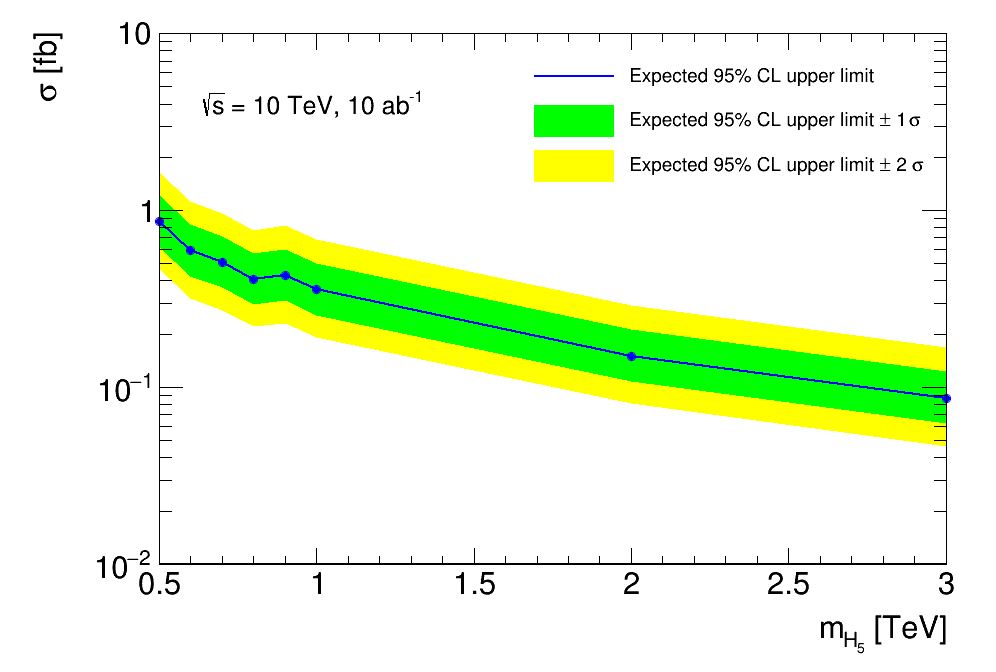} 
\includegraphics[width=0.45\textwidth]{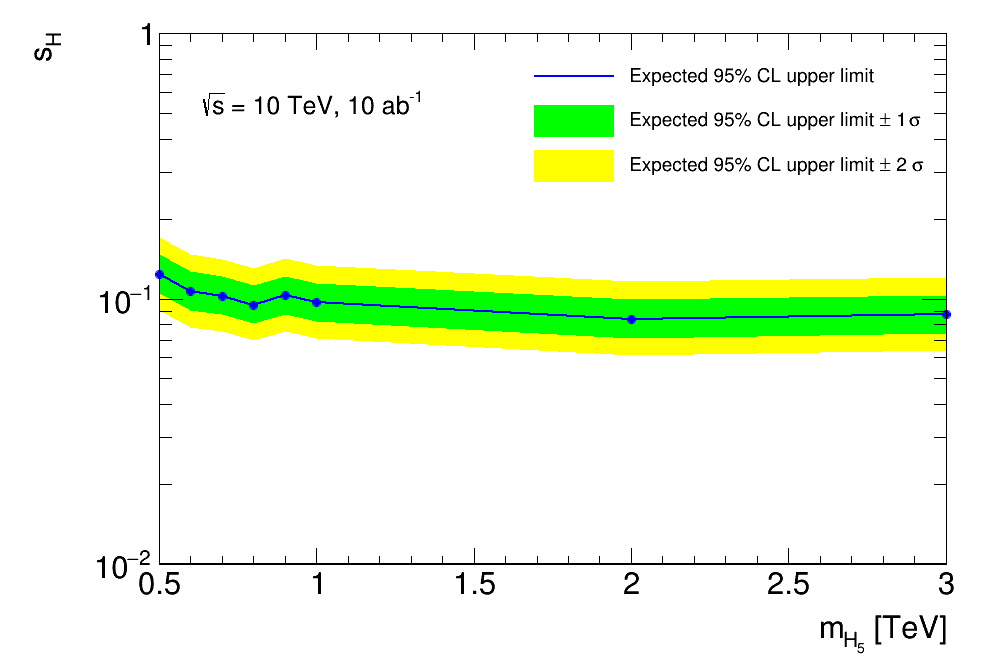}
\includegraphics[width=0.45\textwidth]{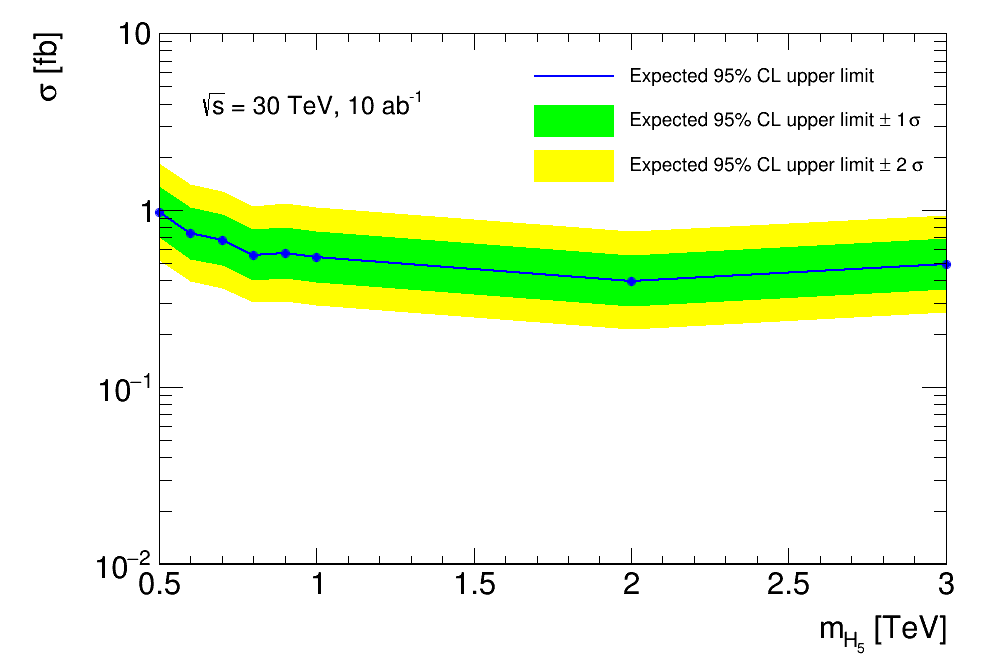} 
\includegraphics[width=0.45\textwidth]{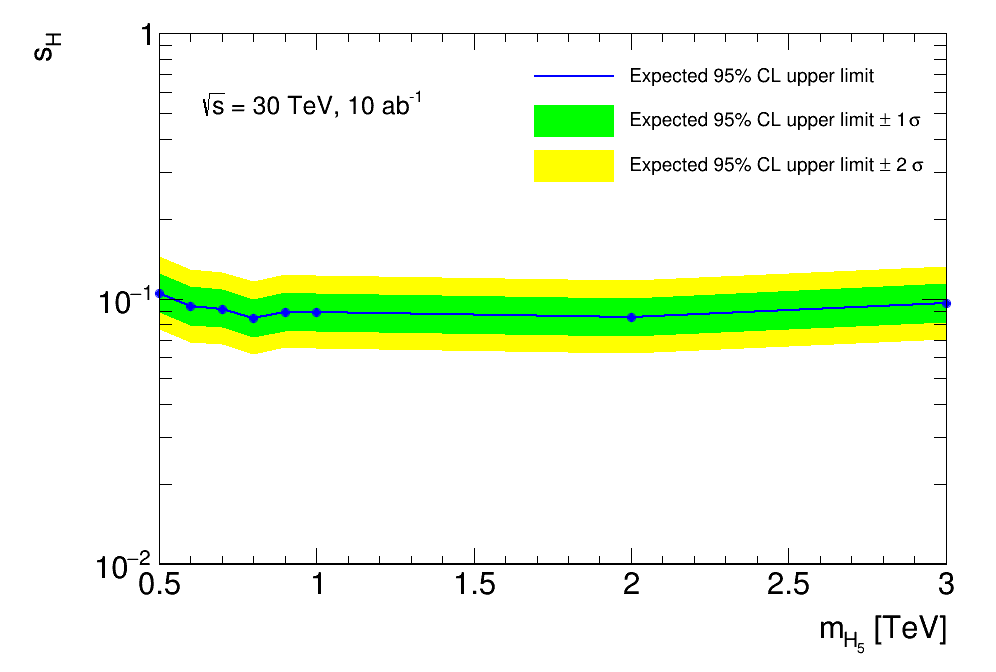}
\caption{Expected exclusion limits at 95\% CL for $\sigma (\PH_{5}\nu\nu) \mathcal{B}(\PH_{5} \to \PV\PV)$ (left) and for $s_{\PH}$ as a function of $m_{\PH_{5}}$ (right) in the GM model for a $\mu^{+}\mu^{-}$ collider with $\sqrt{s}=6$ TeV (upper), 10 TeV (middle), and 30 TeV (lower).\label{fig:gm_limits}}
\end{figure}

The selected events are also used to derive constraints on resonant neutral Higgs boson production in the GM model.  Statistical analysis of the event yields is again performed with a fit to the invariant mass distribution of the leading dijets, which typically correspond to the pair of $\PV$ bosons, denoted $m_{\PV\PV}$. The distributions of $m_{\PV\PV}$ after the event selection are shown in Fig.~\ref{fig:dijet_mass_higgs}. Exclusion intervals are derived using the CLs method~\cite{Junk:1999kv,Read1} in the asymptotic method for the test statistic~\cite{Cowan:2010js}. The exclusion limits on the product of the cross section of neutral Higgs boson production in association with neutrinos and branching fraction to $\PV\PV$, $\sigma (\PH_{5}\nu\nu) \mathcal{B}(\PH_{5} \to \PV\PV)$, at 95\% CL as a function of $m_{\PH_{5}}$ are shown in Fig.~\ref{fig:gm_limits} (left).  
The excluded $s_{\PH_5}$ values at 95\% CL in the GM model as a function of $m_{\PH_{5}}$ are shown in Fig.~\ref{fig:gm_limits} (right). The feature seen in the limit plots at $m_{\PH_5}=0.9$~TeV is the transition point between the $s_{\PH}$ values in the \texttt{H5plane} benchmark~\cite{LHCHiggsCrossSectionWorkingGroup:2016ypw}  and is a consequence of the non-negligible effect of the $\PH_5$ width in the statistical analysis. The reported expected $s_{\PH}$ exclusion values are significantly more stringent compared to the exclusion limits of the current LHC results with more than order an of magnitude better sensitivity for $m_{\PH_{5}}$ values greater than 1 TeV~\cite{PhysRevLett.114.231801,Sirunyan:2017sbn,Sirunyan:2017ret,Sirunyan:2019ksz,Sirunyan:2019der,CMS:2021wlt,ATLAS:2022zuc}.


\section{Summary}
\label{S:summary}

Prospects of searches for anomalous production of heavy gauge boson pairs at future high-energy muon colliders are reported. Muon-muon collision events are simulated at $\sqrt{s}=6$, 10, and 30 TeV corresponding to an integrated luminosity of $4$, $10$, and $10$ ab$^{-1}$, respectively. The simulated events are used to study the $\PW\PW\nu\nu$ and $\PW\PW\mu\mu$ channels with the $\PW$ bosons decaying hadronically. Constraints on the quartic vector boson interactions in the framework of dimension-8 effective field theory operators are obtained with stringent expected limits set on the EFT operators \texttt{S0}, \texttt{S1}, \texttt{M0}, \texttt{M1}, \texttt{M7}, \texttt{T0}, \texttt{T1}, \texttt{T2}, \texttt{T6}, and \texttt{T7}. Depending on the operator, the limits are better by more than one or two orders of magnitude compared to the expected limits at the HL-LHC and HE-LHC. The $\PW\PW\nu\nu$ and $\PZ\PZ\nu\nu$ channels are also used to report expected constraints on the product of the cross section and branching fraction for vector boson fusion production of a heavy neutral Higgs boson as a function of mass from 0.5 to 3 TeV. These results are interpreted in the context of the Georgi--Machacek model and show significantly more stringent constraints compared to the LHC results with more than an order of magnitude better sensitivity for $m_{\PH_{5}}$ values greater than 1 TeV.

\begin{acknowledgments}
We are grateful to D. Zeppenfeld, H. Logan, and W. Yongcheng for fruitful discussions. The work of A.A. is supported by the U.S. Department of Energy, Office of Science, Office of High Energy Physics under contract no. DE-SC0023181.  The work of B.A., V.B., M.K., M.S., J.S., and C.W. is supported by the U.S. Department of Energy, Office of Science, Office of High Energy Physics under contract no. DE-SC0009956. S.C.H., E.K., and A.S. are supported by the National Science Foundation under Grant No. 2110963.
\end{acknowledgments}



\bibliography{muon-vbs.bib}







\end{document}